\documentclass[aps,twocolumn]{revtex4}
\usepackage{psfig}
\usepackage{subfigure}
\usepackage{graphicx}
\usepackage{amssymb}
\usepackage{amsmath}

\begin{document}

\title{Synchronization of dynamical hypernetworks: \\ dimensionality reduction through simultaneous block-diagonalization of matrices}

\author{Daniel Irving, Francesco Sorrentino}
\affiliation{University of New Mexico, Albuquerque, NM, 87131}

\begin{abstract}
We present a general framework to study stability of the synchronous solution for a hypernetwork of coupled dynamical systems. We are able to reduce the dimensionality of the problem by using simultaneous block-diagonalization of matrices. We obtain necessary and sufficient conditions for stability of the synchronous solution in terms of a set of lower-dimensional problems and test the predictions of our low-dimensional analysis through numerical simulations. Under certain conditions, this technique may yield a substantial reduction of the dimensionality of the problem. For example, for a class of dynamical hypernetworks analyzed in the paper, we discover that arbitrarily large networks can be reduced to a collection of subsystems of dimensionality no more than $2$. We apply our reduction techique to a number of different examples, including a class of undirected unweighted hypermotifs of three nodes.
\end{abstract}
\maketitle

\section{Introduction}

Much recent work has been devoted to the study of dynamical networks \cite{Report}. A few studies have considered the dynamics of hypernetworks where the individual units are coupled through two or more interaction networks. 
Hypernetworks arise in applications as different as the spread of epidemic diseases \cite{Ba:Ne}, computer viruses \cite{Wa:Go}, game theory \cite{MNowak}, social interactions \cite{multislice}, and neural networks formed of both electrical gap junctions and chemical synapses \cite{HYP}. In complex adaptive systems, different types of couplings usually coexist, including cooperative, competitive and symbiotic couplings \cite{Millerpage}.

In this paper, we use the term hypernetwork to indicate a set of nodes that are coupled through connections of different types, with the connections of the same type forming a distinct network layer. A similar concept, which has been used to describe mainly social systems, is that of a multislice or multiplex network \cite{multislice}, where 
both intra-layer and inter-layer connections are present. Interdependent networks are usually evoked in the context of engineering or technological applications, when the nodes in each layer rely on their connections to nodes in other layers for their proper functioning (e.g., the coupled functions of the 
power grid and computer communication network studied in \cite{IN}). Transportation networks have been described as layered networks in \cite{kurant2006layered}. Another definition is that of networks of networks, which are evoked in general when connections exist between nodes belonging to different networks (see e.g., \cite{Zh:Ze:Za1,Zh:Ze:Za2}).

We are interested in the synchronization dynamics of  hypernetworks. There are many different types of synchronization, including complete synchronization \cite{FujiYama83,Afraim,Replace}, phase synchronization \cite{Ro:Pi:Ku}, lag synchronization \cite{Ro:Pi:Ku2}, group or cluster synchronization \cite{NSG,clustergroup}, and generalized synchronization \cite{Gen1,Gen2}.  For a review of synchronization of complex networks, the reader is referred to \cite{SReport}.  Synchronization of hypernetworks has relavance to the study of any system exhibiting multiple types of coupling.  For example, studying excitation patterns of neural networks involves the analysis of both chemical and electrical signals between neurons \cite{NSG}. As another example, studying the synchronous motion of entire schools of fish involves the analysis of not only the visual cues between fish but also the release of chemical signals into the water \cite{Pa:Pi,Ab:Po}. In this paper, we focus on complete synchronization of hypernetworks.

The problem of synchronization of dynamical hypernetworks has been first studied in \cite{HYP}, where special conditions have been considered, for which the problem of stability of the synchronous solution can be reduced in a low-dimensional form. However, a general framework to study stability of the synchronous solution for a dynamical hypernetwork is lacking. In this paper, we will present a general approach  to obtain a reduction of this problem in a low-dimensional form. We will do that by looking at lower-dimensional graphs, whose stability will characterize that of the original higher-dimensional network. Moreover, the approach we present can be used to find out to what extent the dimensionality of the original problem can be reduced.

Low-dimensional approaches have proved helpful in analyzing the dynamics of networks of coupled dynamical systems. Examples include (i) the stability of the synchronous evolution for networks of coupled oscillators \cite{FujiYama83,Replace,Pe:Ca,NSG,SAS}, (ii) the stability of the consensus state in networks of coupled integrators \cite{CONS}, (iii) the stability of discrete state models of genetic control \cite{POM}, and (iv) the stability of strategies in networks of coupled agents playing a version of the prisoner's dilemma \cite{GT_SM}. In this paper, we are interested in studying the stability of the synchronous solution for a  dynamical hypernetwork and we show that reducing the stability problem in a lower-dimensional form is an available approach.

\section{Model}

We consider a dynamical hypernetwork described by the following system of coupled differential equations:
\begin{equation}\label{in}
\dot{x}_i(t)=F(x_i(t))+\sum_{k=1}^M \sum_{j=1}^N A_{ij}^{(k)} H_k(x_j(t-\tau_k)),
\end{equation}
$i=1,...N$, where $x_i(t)$ is the $m$-dimensional state of system $i$ at time $t$, $i=1,...,N$, the function $F:R^m \rightarrow R^m$ determines the dynamics of each individual system when uncoupled;  $H_k:R^m \rightarrow R^m$ are arbitrary coupling functions, $k=1,...,M$, $\tau_k\geq 0$ is the time-delay associated with the coupling function $H_k$, $k=1,...,M$. The entries of the matrix $A^{(k)}=\{A_{ij}^{(k)}\}$ are such that $A_{ij}^{(k)} \neq 0$ if node $j$ is coupled to node $i$ through the coupling function $H^k$ and $A_{ij}^{(k)} = 0$ otherwise. Moreover, we require that $\sum_j A_{ij}^{(k)}=a^k$, i.e., the sum of the entries along the rows of each matrix $A^{(k)}$ is constant and is equal to $a^k$ \footnote{Note that the constant-row-sum condition (i.e., that the sum of the rows of the matrix $A(k)$ is constant and equal to $a^k$) is more general than the zero-row-sum condition, usually considered for complete synchronization \cite{FujiYama83,Afraim,Replace}. Even when the constant-row-sum condition is not met, its satisfaction can be dynamically obtained by means of an adaptive
strategy \cite{SOTT,EXP2}.}.  Under this assumption, system (\ref{in}) allows the following synchronous solution,
\begin{equation}\label{ss}
x_1(t)=x_2(t)=...=x_N(t)=x_s(t),
\end{equation}
obeying
\begin{equation} \label{xs}
\dot{x}_s(t)=F(x_s(t))+\sum_k a^k H_k(x_s(t-\tau_k)) \equiv \tilde{F}(x_s(t)).
\end{equation}
Now, by replacing the function $F$ with the function $\tilde{F}$ in (\ref{xs}), we obtain, 
\begin{equation}\label{lin}
\dot{x}_i(t)=\tilde{F}(x_i(t))+\sum_{k=1}^M \sum_{j=1}^N L_{ij}^{(k)} H_k(x_j(t-\tau_k)),
\end{equation}
where each matrix $L^{(k)}=\{L^{(k)}_{ij}\}$, with $L^{(k)}_{ij}={A}^{(k)}_{ij}-\delta_{ij} a^k$ has the property that $\sum_j L_{ij}^{(k)}=0$, $k=1,...,M$, i.e., the sum of the elements in each row is zero, and following a common convention \cite{Report}, we refer to such matrices as Laplacian matrices.

In order to study stability of the synchronous solution, we linearize Eqs. (\ref{lin}) about (\ref{ss}), obtaining,
\begin{equation}
\begin{split}
\delta \dot{x}_i(t)= & D\tilde{F}(x_s(t)) \delta x_i(t)  \\
+  & \sum_{k=1}^M \sum_{j=1}^N L_{ij}^{(k)} DH_k(x_s(t-\tau_k)) \delta x_j(t-\tau_k),
\end{split}
\end{equation}
or equivalently, in matrix form,
\begin{equation}\label{vec}
\begin{split}
\delta \dot{X}(t)= & I_N \otimes D\tilde{F}(x_s(t)) \delta X(t) \\
+ & \sum_{k=1}^M L^{(k)} \otimes DH^k(x_s(t-\tau_k)) \delta X(t-\tau_k)
\end{split}
\end{equation}
where the $mN$-dimensional vector $X(t)=[x_1(t)^T,x_2(t)^T,...,x_N(t)^T]^T$ and we have used the symbol $\otimes$ to indicate the Kronecker product or direct product.

For the sake of simplicity, in what follows, we focus on the case that $M=2$, for which (\ref{vec}) can be rewritten,
\begin{equation}\label{vec2}
\begin{split}
\delta \dot{X}(t)= & I_N \otimes D\tilde{F}(x_s(t)) \delta X(t) \\ +& L^{(1)} \otimes DH_1(x_s(t-\tau_1)) \delta X(t-\tau_1) \\ +& L^{(2)} \otimes DH_2(x_s(t-\tau_2)) \delta X(t-\tau_2).
\end{split}
\end{equation}

Note that (\ref{vec2}) is an $mN$-dimensional system, in the sense that it is described by $mN$ coupled state variables. \footnote{However, if either $\tau_1>0$ or $\tau_2>0$ the number of initial conditions that are needed to describe the evolution of the system can be much larger as for each $i=1,...,N$, knowledge of $x_i(t)$ over a time interval of length $\tau_{max}=\max \tau_i$ is required.}. 
Our goal will be to reduce the dimensionality of the system (\ref{vec2}), by decoupling the stability problem (\ref{vec2}) into a set of lower-dimensional problems, each one independent of the others.

It was shown in Ref. \cite{HYP} that there are three cases for which the $mN$ dimensional problem (\ref{vec2}) can be reduced to a set of $(N-1)$ $2m$-dimensional problems. These three cases are:
(i) the Laplacian matrices $L^{(1)}$ and $L^{(2)}$ commute;
(ii) one of the two networks, say $k=2$, is unweighted and fully connected, i.e., $L^{(2)}_{ij}=c$ for $i\neq j$, $L^{(2)}_{ii}=-c(N-1)$, $i=1,...,N$;
(iii) one of the two networks, say $k=2$,  is such that the coupling strength from node $i$ to node $j$ is a function of $j$ but not of $i$, i.e., $L^{(2)}_{ij}=c_j$ for $i\neq j$, $L^{(2)}_{ii}=-\sum_{j\neq i} c_j$, $i=1,...,N$.

However, if none of the three above conditions is satisfied (and each of the conditions (i),(ii),(iii) generally do not occur if the coupling strengths of the networks are arbitrarily chosen), such a reduction is not possible. In what follows, we will extend the results of \cite{HYP} with the goal of reducing the original stability problem  to a set of $n$ subproblems of maximum dimension $\alpha$, with $2 \leq \alpha \leq N$, depending on the properties of the matrices $L^{(k)}$, $k=1,...,M$.

In general, we will be interested in addressing the following algebraic problem: given the set of $N$-square real matrices $\mathcal{L}=$ $\{L^{(1)}, L^{(2)},...,L^{(M)}\}$, find the finest simultaneous block-diagonalization (SBD) of $\mathcal{L}$. The problem consists in finding an invertible matrix $P$, such that $P^{-1} L^{(i)} P= \oplus_{j=1}^n B_j^i$, where the symbol $\oplus$ denotes the direct sum of matrices, $B_j^1, B_j^2,..., B_j^n$ are square matrix blocks of dimension $b_j$,  and $\sum_{j=1}^n b_j=N$. The diagonalization is said to be the finest if the maximum block dimension $b_{max}=\max_{j=1}^n b_j$ is minimal with respect to the choice of $P$.  The problem can either be solved exactly or in a parameterized (approximate) form, where for a given parameter $\epsilon>0$, $P^{-1} L^{(i)} P= \oplus_{j=1}^n B_j^i+E^{(i)}$, $i=1,...,M$, and the $N$ dimensional matrices $E^{(i)}$, $i=1,...,M$, are of order $\epsilon$ in magnitude.

Hereafter, we briefly review an algorithm for SBD of sets of matrices. Such a block-diagonal decomposition is not unique in general and naturally we are interested in finding the matrix $P$ that provides the finest decomposition.

Instead of trying to tackle the problem directly, the approach in \cite{SBD2},\cite{SBD3} aims at finding a basis that diagonalizes the $*$-algebra associated with the algebra generated by $\mathcal{L}$. This corresponds to finding a matrix $U$ that simultaneously commutes with the matrices $L^{(1)},L^{(2)},...,L^{(M)}$, i.e., that simultaneously satisfies the following sets of equations:
\begin{equation}
U L^{(i)}- L^{(i)} U=0,
\end{equation}
$i=1,...,M$. 

The steps of the algorithm are described in what follows:

(i)  Let $O^{(i)}$ be the $N^2$-matrix $O^{(i)}=I_N \otimes L^{(i)} -L^{(i)} \otimes I_N$.

(ii) Construct the matrix $S=\sum_{i=1}^M {O^{(i)}}^T O^{(i)}$.

(iii) Let $y$ be any $N^2$-vector in the null subspace of the matrix $S$. The $N^2$-vector $u$ can be subdivided in $N$ vectors of dimension $N$ as follows, $u=[u_1^T,u_2^T,...,u_N^T]^T$.

(iv) Obtain $U$ as the matrix whose columns are $u_1$, $u_2$,..., $u_N$.

(v) Finally, $P$ can be constructed as the matrix whose columns are the eigenvectors of $U$.

In certain situations (when for example, a satisfactory SBD reduction is not available), we might be interested in finding a parameterized (approximate) SBD, which can be formulated as follows. Given a parameter $\epsilon>0$, find the invertible $N$-square matrix $P$ such that $P^{-1} L^{(i)} P= \oplus_{j=1}^n B_j^i+E^{(i)}$, $i=1,...,M$, and the $N$ dimensional matrices $E^{(i)}$, $i=1,...,M$, are of order $\epsilon$ in magnitude.
An error controlled version of the SBD algorithm can be found in Ref.\ (16).

Now consider problem (\ref{vec2}). Suppose we have been able to find the finest SBD for $\mathcal{L}=\{L^{(1)}, L^{(2)}\}$ and that this is provided by the invertible matrix $P$. Then we left-multiply both sides of Eq. (\ref{vec2}) by $P^{-1} \otimes I_m$ and by using the change of variables $\eta(t)=P^{-1} \otimes I_m \delta X(t)$, we can rewrite (\ref{vec2}) as follows,
\begin{equation}\label{vec3}
\begin{split}
\dot{\eta}(t) & = I_N \otimes D\tilde{F}(x_s(t)) \eta(t) \\  & + \big( \oplus_{j=1}^n B_j^{(1)} \big) \otimes DH_1(x_s(t-\tau_1)) \eta(t-\tau_1) \\ & + \big( \oplus_{j=1}^n B_j^{(2)} \big) \otimes DH_2(x_s(t-\tau_2)) \eta(t-\tau_2).
\end{split}
\end{equation}

It is easy to see then that the system of equations $(\ref{vec3})$ can be decomposed into the following $n$ subsystems,
\begin{equation}\label{blocks}
\begin{split}
\dot{\eta}_j(t)= & I_{b_j} \otimes D\tilde{F}(x_s(t)) \eta_j(t) + \\ & B_j^{(1)} \otimes DH_1(x_s(t-\tau_1)) \eta_j(t-\tau_1) + \\ & B_j^{(2)} \otimes DH_2(x_s(t-\tau_2)) \eta_j(t-\tau_2),
\end{split}
\end{equation}
$j=1,...,n$, where each vector $\eta_j(t)$ has dimension $m b_j$ and evolves independently from the others.  Moreover, $\sum_j b_j=N$.
Each of these subsystems is forced by the synchronous evolution $x_s(t)$, which obeys Eq. (\ref{xs}). Thus the original $mN$-dimensional problem has been reduced to $n$ lower-dimensional problems, each with dimension $m(b_1+1),m(b_2+1),...,m(b_n+1)$. Also, our proposed goal has been achieved with $\alpha=(b_{max}+1)$, where $2 \leq \alpha \leq N$.
Moreover, this reduction is the finest, in the sense that it is  not possible to obtain another reduction in an $(\alpha-1)$-dimensional form, provided that the original block-diagonalization was the finest.

We note that by construction,  the matrices $L^{(k)}$, $k=1,...,M$, have a zero eigenvalue with associated eigenvector $[1,1,...,1]^T$. Hence, when obtaining the finest SBD, there must be a $1$-dimensional subsystem, indexed $j=1$, for which $B_1^{(k)}=0$, $k=1,...,M$, yielding, 
\begin{equation}\label{block1}
\dot{\eta}_1(t)=  D\tilde{F}(x_s(t)) \eta_1(t).
\end{equation}
We note that this one equation is associated with perturbations lying in the direction parallel to the synchronization manifold (given by the eigenvector $[1,1,...,1]^T$) and therefore it is irrelevant in determining transversal stability of the synchronous solution.

For a generic subsystem of dimension $D$, we may want to identify a minimal set of parameters $(p_1,p_2,...,p_{r_D})$ that characterize stability. In general, it can be shown that the minimum number of parameters is $r_D=D^2+1$ (see Sec. IIA). However, for some specific cases,  it is possible to parameterize a $D$-dimensional subsystem by using less than $r_D$ parameters.

\subsection{The particular case that $b_{max}=2$}


We now look at Eq.\ \ref{blocks} and consider the particular case that $b_{max}=2$, i.e., $\alpha=3$. We further assume that the simultaneous block-diagonalization yields $\mu$ blocks of dimension $1$ and $\nu$ blocks of dimension $2$, with $(\mu+2 \nu)=(N-1)$. The problem that we want to address in this section is described below.

Consider all the pairs of matrices $\mathcal{L}=\{L^{(1)}, L^{(2)}\}$ such that a simultaneous block-diagonalization can be achieved with $b_{max}=2$. For blocks of dimension $D=\{1,2\}$, we aim at finding a reduction of the stability problem in a parametric form in the $r_D$ scalar parameters $(p_1,p_2,...,p_{r_D})$, such that $r_D$ is minimal. In what follows, we independently address this problem for blocks of dimension $D=1$ and blocks of dimension $D=2$ and we show that $r_1=2$ and $r_2=5$.  We also obtain a general relation between $r_D$ and the block dimension $D$.

It is easy to see that for blocks of dimension $1$, Eq. (\ref{blocks}) becomes,
\begin{equation}\label{uno}
\begin{split}
\dot{\eta}_j(t)= & D\tilde{F}(x_s(t)) \eta_j(t) \\ + & B_j^{(1)}  DH_1(x_s(t-\tau_1)) \eta_j(t-\tau_1) \\ + & B_j^{(2)}  DH_2(x_s(t-\tau_2)) \eta_j(t-\tau_2)
\end{split}
\end{equation}
$j=1,...,\mu$, where $\eta_j(t)$ has dimension $m$ and $B_j^{(1)}$ and $B_j^{(2)}$ are two scalar (eventually, complex) parameters. Each subsystem (\ref{uno}) is parametrized by the pair $(B_j^{(1)},B_j^{(2)})$. Hence, $r_1=2$.

For blocks of dimension $2$, Eq. (\ref{blocks}) becomes,
\begin{equation}\label{due}
\begin{split}
\dot{\eta}_j(t)= & I_{2} \otimes D\tilde{F}(x_s(t)) \eta_j(t) \\ + & B_j^{(1)} \otimes DH_1(x_s(t-\tau_1)) \eta_j(t-\tau_1) \\ + & B_j^{(2)} \otimes DH_2(x_s(t-\tau_2)) \eta_j(t-\tau_2)
\end{split} \end{equation}
$j=1,...,\nu$, where $\eta_j(t)$ has dimension $2m$ and $B_j^{(1)}$ and $B_j^{(2)}$ are two square matrices of dimension $2$. We note that it is possible to further diagonalize either one of the two matrices $B_j^{(1)}$ or $B_j^{(2)}$; without loss of generality, we diagonalize  $B_j^{(1)}$, obtaining $B_j^{(1)}=W_j \Lambda_j W^{-1}_j$. By pre-multiplying  each block (\ref{due}) by $W_j^{-1} \otimes I_m$, we obtain,
\begin{equation}\label{due2}
\begin{split}
\dot{\zeta}_j(t)= & I_{2} \otimes D\tilde{F}(x_s(t)) \zeta_j(t) \\ + & \Lambda_j \otimes DH_1(x_s(t-\tau_1)) \zeta_j(t-\tau_1) \\ + & Q_j \otimes DH_2(x_s(t-\tau_2)) \zeta_j(t-\tau_2),
\end{split}
\end{equation}
$j=1,...,\nu$, where $\zeta_j(t)=W_j^{-1} \otimes I_m \eta_j(t)$, $\Lambda_j$ is the following diagonal matrix,
\begin{equation}\label{LA}
\Lambda_j=\left[\begin{array}{cc}
    \lambda_j^1 & 0  \\
    0 & \lambda_j^2
  \end{array} \right]
\end{equation}
and the matrix $Q_j=W^{-1}_j  B_j^{(1)} W_j$ is the following,
\begin{equation}\label{Q}
Q_j=\left[\begin{array}{cc}
    q_j^{11} & q_j^{12}   \\
    q_j^{21} & q_j^{22}
  \end{array} \right].
\end{equation}
We see that for each block $j=1,...,\nu$, stability depends on the following set of scalar parameters $(\lambda_j^1,\lambda_j^2,q_j^{11},q_j^{12},q_j^{21},q_j^{22})$. From (\ref{due},\ref{LA},\ref{Q}) we see that Eq. (\ref{due2}) can be decomposed into the following two coupled equations,
\begin{equation}\label{due3}
\begin{split}
\dot{\zeta}_j^1(t)= &  D\tilde{F}(x_s(t)) \zeta_j^1(t) +\lambda_j^1  DH_1(x_s(t-\tau_1)) \zeta_j^1(t-\tau_1) \\ +  & q_j^{11} DH_2(x_s(t-\tau_2)) \zeta_j^1(t-\tau_2) \\ + & q_j^{12} DH_2(x_s(t-\tau_2)) \zeta_j^2(t-\tau_2), \\
\dot{\zeta}_j^2(t)= &  D\tilde{F}(x_s(t)) \zeta_j^2(t) +\lambda_j^2  DH_1(x_s(t-\tau_1)) \zeta_j^2(t-\tau_1) \\ + & q_j^{21} DH_2(x_s(t-\tau_2)) \zeta_j^1(t-\tau_2) \\ + & q_j^{22} DH_2(x_s(t-\tau_2)) \zeta_j^2(t-\tau_2),
\end{split}
\end{equation}
 where the vector $\zeta_j=[{\zeta_j^1}^T(t),{\zeta_j^2}^T(t)]^T$. Now, with the substitution, $q_j^{12} \zeta_j^2(t) \rightarrow \zeta_j^2(t)$, we can rewrite (\ref{due3}),
 \begin{equation}\label{due4}
 \begin{split}
\dot{\zeta}_j^1(t)= &   D\tilde{F}(x_s(t)) \zeta_j^1(t) +\lambda_j^1  DH_1(x_s(t-\tau_1)) \zeta_j^1(t-\tau_1) \\ + & q_j^{11} DH_2(x_s(t-\tau_2)) \zeta_j^1(t-\tau_2) \\+ &  DH_2(x_s(t-\tau_2)) \zeta_j^2(t-\tau_2),  \\
\dot{\zeta}_j^2(t)=  & D\tilde{F}(x_s(t)) \zeta_j^2(t) +\lambda_j^2  DH_1(x_s(t-\tau_1)) \zeta_j^2(t-\tau_1) \\ + &  q_j^{12} q_j^{21} DH_2(x_s(t-\tau_2)) \zeta_j^1(t-\tau_2) \\+ & q_j^{22} DH_2(x_s(t-\tau_2)) \zeta_j^2(t-\tau_2).
\end{split}
\end{equation}
Thus each subsystem (\ref{due4}), $j=1,...,\nu$, is described by the following set of $r_2=5$ scalar parameters $(\lambda_j^1,\lambda_j^2,q_j^{11},q_j^{12}q_j^{21},q_j^{22})$. It follows that for $2$-dimensional subsystems, $r_2=5$.

We conclude that each one-dimensional subsystem $j=1,...,\mu$ can be associated with a master stability function $\mathcal{M}^1(B_j^{(1)},B_j^{(2)})$
which returns the maximum Lyapunov exponent of Eq. (\ref{uno}) as a function of the pair $(B_j^{(1)},B_j^{(2)})$. Also, each two-dimensional subsystem $j=1,...,\nu$ can be associated with a master stability function $\mathcal{M}^2(\lambda_j^1,\lambda_j^2,q_j^{11},q_j^{12}q_j^{21},q_j^{22})$ which returns the maximum Lyapunov exponent of Eq. (\ref{due3}) as a function of the $5$-tuple $(\lambda_j^1,\lambda_j^2,q_j^{11},q_j^{12}q_j^{21},q_j^{22})$.
Once the  master stability functions $\mathcal{M}^1, \mathcal{M}^2$ are known, stability of the synchronous solution for a generic dynamical hypernetwork (described by Eq. (1) with $M=2$) that allows a simultaneous block-diagonalization with $b_{max}=2$, can be determined by knowledge of the pairs $(B_j^{(1)},B_j^{(2)})$ for blocks of dimension $1$ and of the $5$-tuples $(\lambda_j^1,\lambda_j^2,q_j^{11},q_j^{12}q_j^{21},q_j^{22})$ for blocks of dimension 2.

By extending the above reasoning to subsystems of higher dimension $D$, it can be shown that $r_D=D^2+1$.

\section{Numerical examples}

\begin{figure}
\centering
\includegraphics[width=3in]{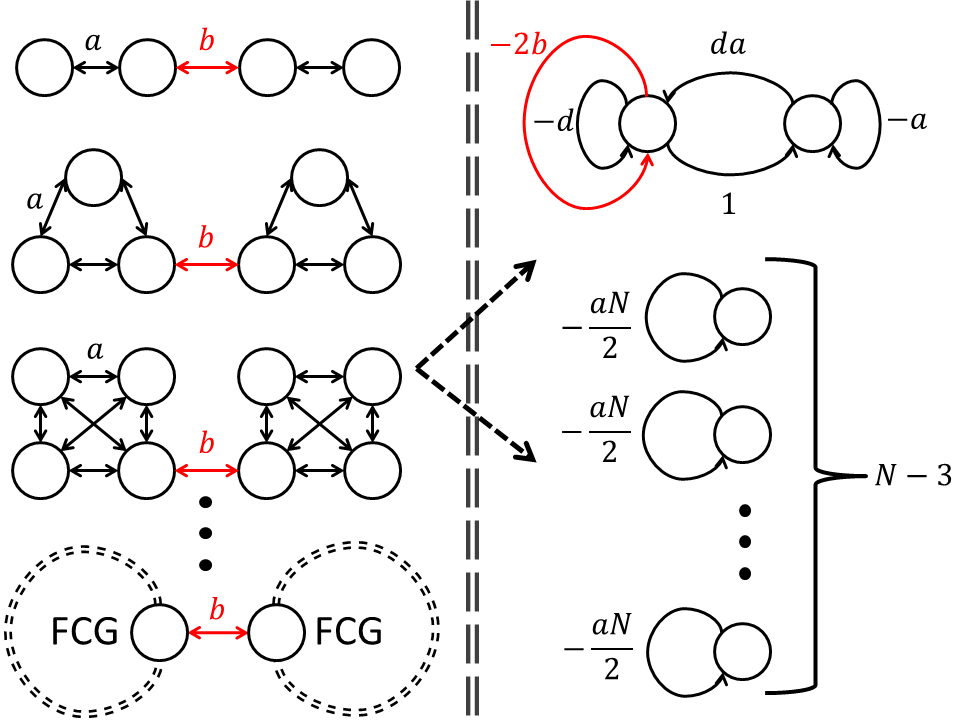}
\caption{[Color Online] A special class of hypernetwork configurations. These hypernetworks shown on the left can be neatly reduced into the $(N-2)$ subsystems shown on the right.  The reduction yields a single two-node system, followed by $(N-3)$ one-node systems.  Stability of the hypernetworks on the left corresponds to that of the lower-dimensional subsystems on the right.}
\label{Nnode}
\end{figure}

The left hand side of Fig. \ref{Nnode} shows a special class of hypernetworks for which the stability problem can be conveniently reduced by using SBD. 
The class contains all hypernetworks made from two identical fully connected graphs (FCG), each of size $\frac{N}{2}$, that are connected to one another only by a single alternative connection.  The parameter $a$ is the coupling strength of all the connections inside each FCG and the parameter $b$ is the coupling strengths of the alternative connection. The associated Laplacian matrices do not commute unless either $a=0$ or $b=0$. As can be seen on the right hand side of Fig. \ref{Nnode}, we discover that a hypernetwork in this configuration can always be reduced to a collection of subsystems of dimensionality no more than $2$, that is, one subsystem of dimension $2$ and $(N-3)$ identical subsystems of dimension $1$ (we neglect the one subsystem that is associated with perturbations parallel to the synchronization manifold). The reduction in this form becomes particularly advantageous when  the dimension of the original hypernetwork $N$ is large. From the SBD decomposition (described in Sec.\ II), we obtain that the parameter $d$ in the figure depends upon the size of the hypernetwork, i.e,  $d=a(\frac{N}{2}-1)$.  

As an example, we consider the hypernetwork shown on the left hand side of Fig. \ref{Nnode} of dimension $N=6$. 
Its dynamics is described by the set of Eqs. (\ref{in}), where each individual node obeys the equation of the Lorenz chaotic system, for which $m=3$, $x(t)=(x_1(t),x_2(t),x_3(t))^T$,
\begin{equation}\label{LOR}
F({{x}})=\left[\begin{array}{c}
    10[x_2(t)-x_1(t)]  \\
    x_{1}(t)[28- x_{3}(t)] -x_2(t) \\
    x_{1}(t) x_2(t) - 2 x_3(t)
  \end{array} \right],
\end{equation}
$H_1(x(t))=[0,x_2(t),0]^T$, $H_2(x(t))=[x_1(t), 0,x_3(t)]^T$,
and $\tau_1=\tau_2=0$. The adjacency matrices $A^{(1)}$  and $A^{(2)}$ correspond respectively to the black [black] and gray [red] connections of the $N=6$ hypernetwork in Fig. \ref{Nnode} and are defined as follows: $A_{ij}^{(1)}=A_{ji}^{(1)}=a$ if $(i-\theta)\times(j-\theta)>0$, $0$ otherwise, where $\theta=\frac{N+1}{2}$;  $A_{ij}^{(2)}=A_{ji}^{(2)}=0$ except for the one pair $(i=i^*,j=j^*)$, with $1\leq i^* \leq \frac{N}{2}$ and $\frac{N}{2}+1\leq j^* \leq N$. 
$L^{(k)}_{ij}=({A}^{(k)}_{ij}- \delta_{ij} \sum_\ell {A}^{(k)}_{i\ell})$, $k=\{1,2\}$.  

\begin{figure}[t!]
\centering
\includegraphics[width=1.9in]{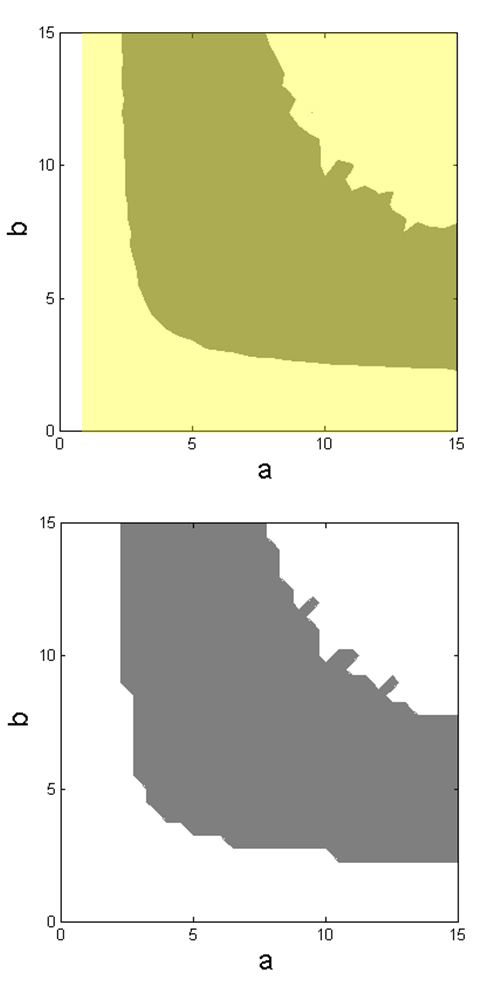}
\caption{[Color Online] The upper plot shows the areas of the $(a,b)$ plane corresponding to a negative MLE for both the $m$-dimensional subsystem (colored in light gray [yellow]) and the $2m$-dimensional subsystems (colored in dark gray [gray]). We expect stability of the original hypernetwork in the intersection of the light gray [yellow] and dark gray [gray] areas. The lower plot shows simulations of the full high-dimensional hypernetwork, $N=6$; we plot in gray the area of the $(a,b)$-plane for which the synchronization error $E(t)$ decreases steadily below $1\%$ of its initial value.}
\label{Compare}
\end{figure}

From the SBD procedure, we obtain that stability of the original $Nm$-dimensional system can be reduced to that of two lower-dimensional systems, one of dimension $m$ and one of dimension $2m$ (see Fig. \ref{Nnode}). We indepently compute the maximum Lyapunov exponent (MLE) for both these systems. For the $m$-dimensional subsystem, we find that the condition for stability is that $a\frac{N}{2}>2.29$. For the $2m$-dimensional subsystem, we record the maximum Lyapunov exponent as a function of the pair $(a,b)$, for the specific case of $N=6$. 
The results of our numerical computations are summarized in the upper plot of Fig. \ref{Compare}, where the light gray [yellow] area  corresponds to the region of the $(a,b)$ plane for which the MLE of the $m$-dimensional subsystem is negative and the dark gray [gray] area corresponds to the region of $(a,b)$ plane for which the MLE of the $2m$-dimensional system is negative. Note that for this case,  the intersection coincides with the gray area. 


In order to test our low-dimensional predictions, we numerically integrate Eqs. (\ref{in})  from an initial condition close to the synchronization manifold. 
 For each run, we monitor the
average synchronization error $E$,
\begin{equation} 
{E}(t)=(N \Delta t)^{-1} { \sum_{i=1}^N \int_t^{t+\Delta t} {\left\|x_{i}(\tau)-\bar{x}(\tau)\right\|}} d \tau,\label{error}
\end{equation}
where $\bar x(t)=N^{-1} \sum_{i=1}^{N} x_{i}(t)$ and $\left\| \xi \right\|$ indicates the Euclidean norm of the vector $\xi$.  The lower plot of Fig. \ref{Compare} shows the area of the $(a,b)$-plane for which $E(t)$ is observed to decrease steadily below $1\%$ of its initial value. As can be seen from the upper and lower plots of Fig. \ref{Compare}, there is very good agreement between the dynamics of the original hypernetwork and its low-dimensional counterpart.

\begin{figure}[h!]
\centering
\includegraphics[width=3.5in]{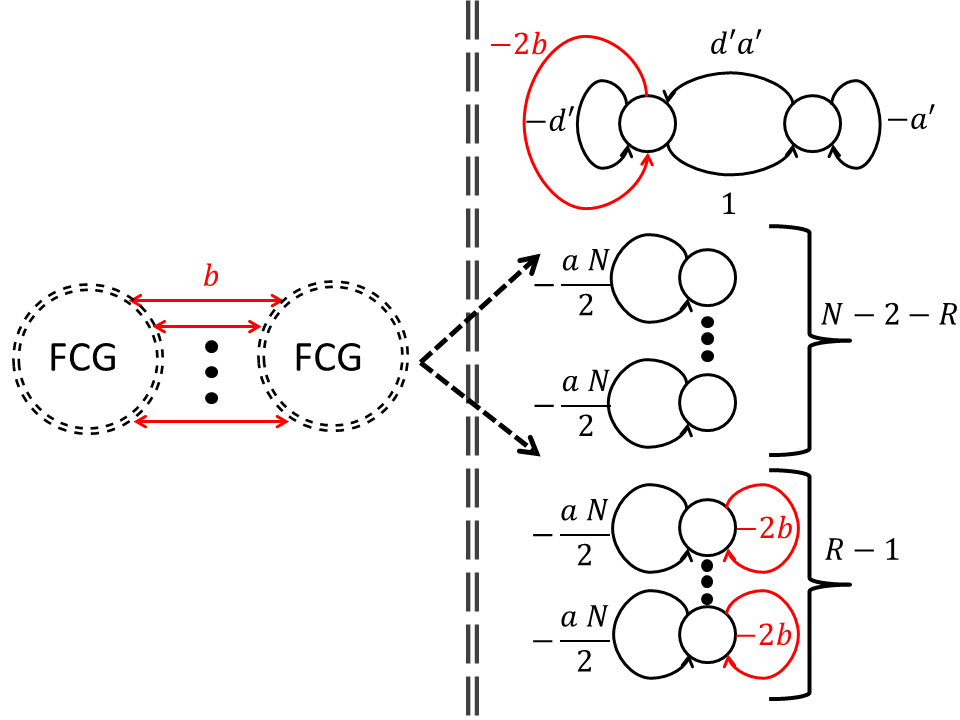}
\caption{[Color Online] On the left: a hypernetwork formed of two identical fully connected graphs, each of size $\frac{N}{2}$,  connected to one another by a set of $R$ alternative connections, $1 \leq R<N/2$. The endpoints of the alternative connections never coincide in the same node. The case that $R=1$ corresponds to that studied above in Fig. \ref{Nnode}.  These hypernetworks shown on the left can be neatly reduced into the $(N-2)$ subsystems shown on the right.  The reduction yields a single two-node system, followed by $(N-3)$ one-node systems, which are divided into two distinct groups of $(R-1)$ and $(N-2-R)$ identical subsystems. Stability of the hypernetworks on the left corresponds to that of the lower-dimensional subsystems on the right.}
\label{Nnode2}
\end{figure}

We also considered the case shown in Fig.\ref{Nnode2} that the two FCG graphs are connected by $R$ alternative connections rather than $1$, each one with associated strength $b$, and such that the endpoints of these $R$ connections never coincide in the same node. How is the stability of this  hypernetwork going to be characterized?  In what follows, we restrict our attention to the case that $1 \leq R< N/2$. By applying the SBD procedure, we discover that, as can be seen from Fig. \ref{Nnode2},  a hypernetwork in this configuration can always be reduced to a collection of subsystems of dimensionality no more than $2$, that is, one subsystem of dimension $2$ and $(N-3)$ subsystems of dimension $1$ (we neglect the one subsystem that is associated with perturbations parallel to the synchronization manifold).  The one subsystem of dimension $2$ depends on the number of alternative connections $R$, with the parameters shown in the figure $d'=a(\frac{N}{2}-R)$ and $a'=aR$.
Moreover, for this more general case, as can be seen on the right hand side of Fig. \ref{Nnode2},  the remaining $(N-3)$ subsystems of dimension $1$ are divided in two distinct groups of $(R-1)$ and $(N-2-R)$ identical subsystems. Then, in order to characterize stability of hypernetworks in this configuration, we have to consider stability of all  the three different types of subsystems that arise from the reduction. Stability would occur in a region in the parameter space which is the intersection of the three resulting stability regions. The reduction would still be very significant for large enough values of $N$. 

\begin{figure}
\centering
\includegraphics[width=3in]{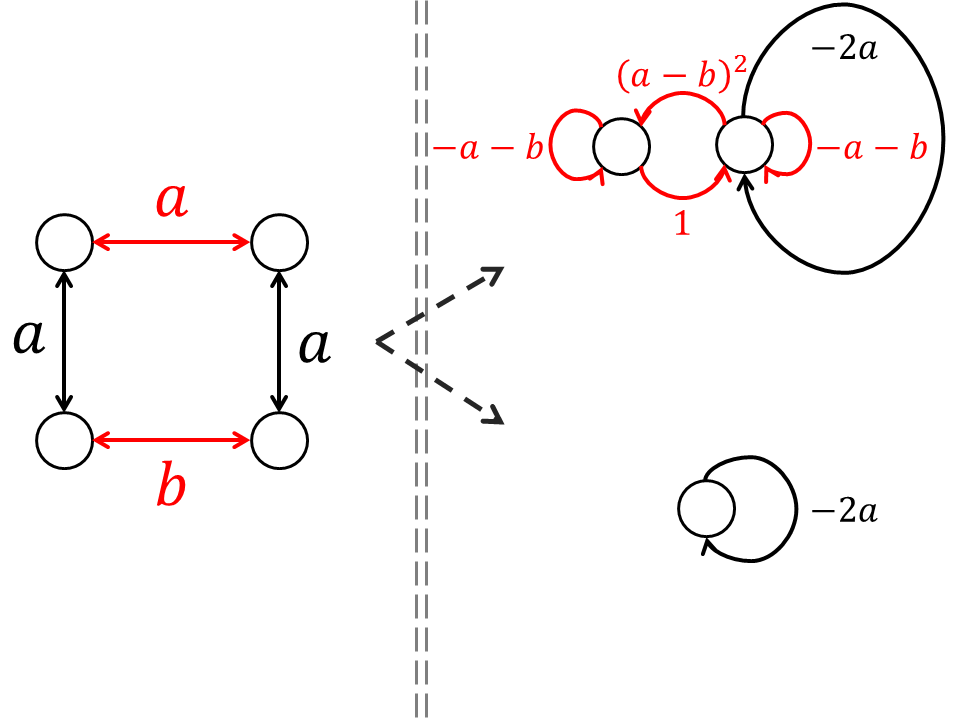}
\caption{[Color Online] On the left, a hypernetwork formed of two interaction graphs. Each node represents a linearized system. On the right, we have obtained a reduction into two lower dimensional systems: one of $b_1=2$ nodes and another one formed of $b_2=1$ node. }
\label{Grafo4}
\end{figure}

As another example, we consider the  hypernetwork shown on the left hand side of Fig. \ref{Grafo4}. From the figure we see that the hypernetwork is composed of $M=2$ different networks, each one associated with a different coupling function $H_k$ (to be defined in what follows). Hence, the linearized problem can be cast exactly in the form of Eq. (7) with $M=2$, where the two Laplacian matrices are as follows,

\begin{eqnarray}L^{(1)}= \{L^{(1)}_{ij} \}=
{\small\small\small{ \begin{pmatrix}
-a & 0 & a & 0   \cr
0 & -a & 0 & a \cr
a & 0 &  -a & 0 \cr
0 & a  & 0 & -a
\end{pmatrix}}}, \label{LC1}
\end{eqnarray}
 associated with the black [black] connections in the figure and
 \begin{eqnarray}L^{(2)}= \{L^{(2)}_{ij} \}=
{\small\small\small{ \begin{pmatrix} -a &
a  & 0 & 0 \cr
a & -a & 0 & 0   \cr
0 & 0 & -b  & b \cr
0 & 0 & b & -b
\end{pmatrix}}}, \label{LC2}
\end{eqnarray}
 associated with the gray [red] connections in the figure.

The two matrices do not commute unless $b=a$. Therefore, we consider the case $b\neq a$. The dynamical hypernetwork is described by the set of Eqs. (\ref{in}), where each individual node obeys the equation of the Lorenz chaotic system, for which $m=3$, $x(t)=(x_1(t),x_2(t),x_3(t))^T$,
\begin{equation}\label{LOR}
F({{x}})=\left[\begin{array}{c}
    10[x_2(t)-x_1(t)]  \\
    x_{1}(t)[28- x_{3}(t)] -x_2(t) \\
    x_{1}(t) x_2(t) - \frac{8}{3} x_3(t)
  \end{array} \right],
\end{equation}
$H_1(x(t))=[x_1(t),0,0]^T$, $H_2(x(t))=[0, 0,x_3(t)]^T$,
and $\tau_1=\tau_2=0$.

By using the procedure described in \cite{SBD2}, we find a matrix $P$,
 \begin{equation}
P=\left[\begin{array}{cccc}
    -0.5000 & -0.5000 & -0.4330  & -0.5590 \\
    -0.5000 & -0.5000 &  0.4330  &  0.5590 \\
    -0.5000 &  0.5000 & -0.5590  &  0.4330 \\
    -0.5000 &  0.5000 &  0.5590  & -0.4330
  \end{array} \right],
\end{equation}
that simultaneously block-diagonalizes $L^{(1)}$ and $L^{(2)}$. 
 By using $P$, we obtain that the original $4m$-dimensional system can be decomposed in two $m$-dimensional systems in the blocks $(0,0)$ and $(-2a,0)$ and in one $2m$-dimensional system. 
 The subsystem associated with the pair $(0,0)$ corresponds to perturbations parallel to the synchronization manifold and as such is irrelevant in determining transversal stability of the synchronous solution. 


\begin{figure}
\centering
\includegraphics[width=4in]{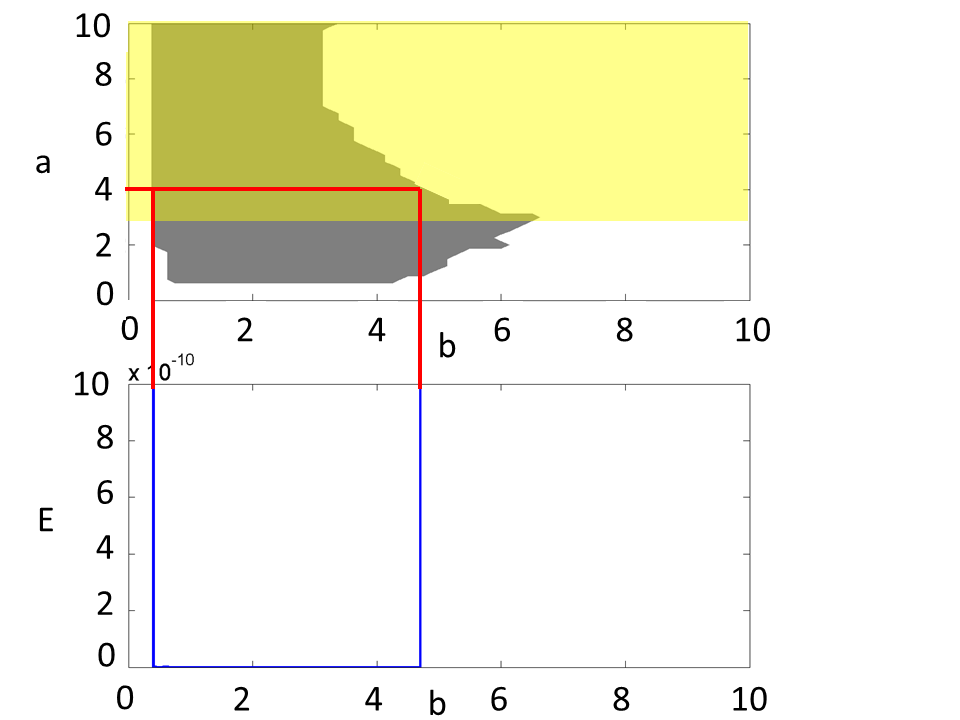}
\caption{[Color Online] The upper plot shows the areas of the $(a,b)$ plane to which corresponds a negative maximum Lyapunov exponent (MLE) for both the $m$-dimensional subsystem on the right-hand side of Fig. \ref{Grafo4} (colored in light gray [yellow]) and the $2m$-dimensional subsystems on the right-hand side of Fig. \ref{Grafo4} (colored in dark gray [gray]). We expect stability of the original hypernetwork in the intersection of the light gray [yellow] and dark gray [gray] areas. The lower plot shows the synchronization error $E$ versus $b$ for $a=4$, which converges to zero in the stability interval predicted by the lower-dimensional analysis. }
\label{Grafo}
\end{figure}

By further diagonalizing the one $2m$-dimensional subsystem, we obtain that this can be recast into the form,
\begin{equation}\label{form}
\begin{split}
\dot{\zeta}_j(t)= & I_{2} \otimes D\tilde{F}(x_s(t)) \zeta_j(t) \\ + & \Lambda_j \otimes DH_1(x_s(t-\tau_1)) \zeta_j(t-\tau_1) \\ + & Q_j \otimes DH_2(x_s(t-\tau_2)) \zeta_j(t-\tau_2),
\end{split}
\end{equation}
with
\begin{equation}\label{LA2}
\Lambda_j=\left[\begin{array}{cc}
    0 & 0  \\
    0 & - 2 a
  \end{array} \right]
\end{equation}
and
\begin{equation}\label{Q2}
Q_j=\left[\begin{array}{cc}
    -(a+b) & 1   \\
    (b-a)^2 & -(a+b)
  \end{array} \right].
\end{equation}
The procedure to obtain Eq.\ (\ref{form}) is illustrated in Sec.\ IIA, compare with Eq.\ (\ref{due2}) therein.
The right hand side of Fig. \ref{Grafo4}  shows the two lower-dimensional subsystems in which the stability  problem has been reduced. We observe that for this specific problem, stability of both the $m$ and $2m$-dimensional subsystems can be conveniently parameterized in the pair $(a,b)$. The upper plot of Fig. \ref{Grafo} shows the sign of the maximum Lyapunov exponent (MLE) associated with both the $m$ and $2m$-dimensional subsystems in the $(a,b)$ plane. The area associated with a negative MLE for the $m$-dimensional subsystem is colored in light gray [yellow] and  the area associated with a negative MLE for the $2m$-dimensional subsystem is colored in dark gray [gray]. We expect stability of the original hypernetwork in the intersection of the light gray [yellow] and dark gray [gray] areas in the f.


In order to test our predictions, we numerically integrate from an initial condition close to the synchronization manifold the equations of the dynamical hypernetwork (1), with $M=2$, the function $F$ given in (\ref{LOR}), $H_1(x(t))=[x_1(t),0,0]^T$, $H_2(x(t))=[0, 0, x_3(t)]^T$,
 $\tau_1=\tau_2=0$, and the two Laplacian matrices $L^{(1)}$ and $L^{(2)}$ given in Eqs. (\ref{LC1}) and (\ref{LC2}). For each run, we monitor the
average synchronization error $E$, defined in Eq. (\ref{error}).  The lower plot of Fig. \ref{Grafo} shows the final synchronization error $E$ versus $b$ for $a=4$, which converges to zero in the stability interval predicted by the lower-dimensional analysis.

\subsection{Dynamical Hypermotifs}

As a further application of our theory, we have considered the synchronization of dynamical hypermotifs. Motifs were introduced in Ref.  \cite{milo2002network} as recurrent patterns of interconnections occurring in complex networks.   Synchronization of small network motifs has been studied in \cite{lodato2007synchronization,d2008synchronization}. Here, we are interested in hypermotifs, i.e. motifs with multiple types of coupling. 

In particular, we have considered the class of all the possible $N=3$-node unweighted undirected hypermotifs with $M=2$ connection types. We have assumed all the connections to have unitary weights. We have obtained a list of $6$ different such hypermotifs, excluding those for which $A^{(1)}=A^{(2)}$ and those obtained from one of the motifs in the list by interchanging the  matrix $A^{(1)}$ with  the matrix $A^{(2)}$. 

Figure 6 shows all of these $6$ hypermotifs, labeled as A-F. We have found that for the hypermotifs D-F the Laplacian matrices associated with $A^{(1)}$ and $A^{(2)}$ commute. Instead, for the hypermotifs A-C, we have applied the SBD procedure to reduce them in their lower-dimensional form. Their lower-dimensional counterparts are also shown in  Fig.\ 6 (center column). The study of more elaborated hypermotifs (e.g., with direct connections or with more than $3$ nodes) is beyond the scope of this paper.

\section{Conclusions}

In this paper, we introduced a general framework to study stability of the synchronous solution of a dynamical hypernetwork by means of a dimensionality reduction strategy. For any set of arbitrarily chosen coupling matrices, we are able to obtain the finest SBD (simultaneous block diagonalization) and to evaluate stability of the synchronous solution based on that. Under certain conditions, this technique may yield a substantial reduction of the dimensionality of the problem. For example, for a class of dynamical hypernetworks analyzed in this paper, we discovered that arbitrarily large networks can be reduced to a collection of subsystems of dimensionality no more than $2$. Other times the reduction may be less significant. 

We have applied our reduction techique to a number of different examples, including small undirected unweighted hypermotifs of $3$ nodes. An important advantage of the SBD decomposition is that it can be used to find out to what extent the dimensionality of the original problem can be reduced. The study of synchronization of large arbitrary dynamical hypernetworks is the subject of ongoing investigations.


\begin{widetext}

\begin{figure}
\centering
\includegraphics[width=6in,height=4in]{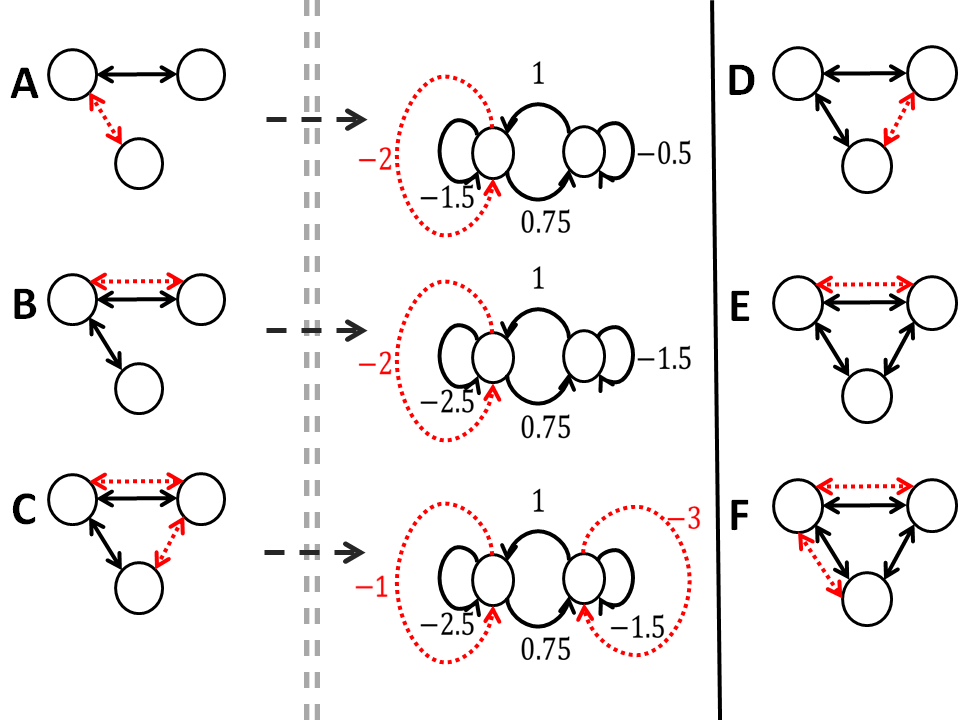}
\caption{[Color Online]  A-F are all the possible unweighted undirected hypermotifs with N=3 nodes and M=2 connection types, excluding those for which $A^{(1)}=A^{(2)}$ and those obtained from A-F by interchanging the  matrix $A^{(1)}$ with the matrix $A^{(2)}$. All the connections in A-F have associated unitary weights. For the hypermotifs A-C we use the SBD procedure to reduce them in their  lower-dimensional form (the lower dimensional graphs are those in the center column). For hypermotifs D-F, the Laplacian matrices corresponding to $A^{(1)}$ and $A^{(2)}$ commute. Hence, their dynamical reduction is not shown. }
\label{HM}
\end{figure}
\end{widetext}

The authors are indebted to Jens Lorenz for insightful discussions.

\bibliography{biblioRidond2s}

\end{document}